\documentclass[a4paper]{article}
\pdfoutput=1

\usepackage[english]{babel}
\usepackage[utf8]{inputenc}
\usepackage{amssymb}
\usepackage{verbatim}
\usepackage{amsmath}
\usepackage{amsthm}
\usepackage{amsfonts}
\usepackage{graphicx}
\usepackage{url}
\usepackage{listings}
\usepackage[official]{eurosym}
\usepackage{array}
\usepackage{booktabs}
\usepackage{tikz}
\usetikzlibrary{shapes,arrows, shapes.geometric, shapes.symbols}
\tikzstyle{every label}= [black]
\tikzstyle{place}=[circle,draw=black,minimum size=6mm, node distance=2cm]
\tikzstyle{transition}=[rectangle,draw=black, minimum size=6mm, node distance=2cm]
\tikzstyle{pre}=[<-,shorten <=1pt,>=stealth',semithick]
\tikzstyle{post}=[->,shorten >=1pt,>=stealth',semithick]

\usepackage{authblk}

\usepackage{listings}

\usepackage{color}
 
\definecolor{codegreen}{rgb}{0,0.6,0}
\definecolor{codegray}{rgb}{0.5,0.5,0.5}
\definecolor{codepurple}{rgb}{0.58,0,0.82}
\definecolor{backcolour}{rgb}{0.95,0.95,0.92}
 
\lstdefinestyle{mystyle}{
    backgroundcolor=\color{backcolour},   
    commentstyle=\color{codegreen},
    keywordstyle=\color{magenta},
    numberstyle=\tiny\color{codegray},
    stringstyle=\color{codepurple},
    basicstyle=\footnotesize,
    breakatwhitespace=false,         
    breaklines=true,                 
    captionpos=b,                    
    keepspaces=true,                 
    numbers=left,                    
    numbersep=5pt,                  
    showspaces=false,                
    showstringspaces=false,
    showtabs=false,                  
    tabsize=2,
}

\usepackage{natbib}

\begin{document}

\title{PyCaMa: Python for cash management}

\author[1]{Francisco Salas-Molina\footnote{Corresponding author. E-mail addresses: \textit{francisco.salas@hifesa.com, jar@iiia.csic.es, pdizga@txp.upv.es}}}
\author[2]{Juan A. Rodr\'iguez-Aguilar}
\author[3]{Pablo D\'iaz-Garc\'ia}

\affil[1]{Hilaturas Ferre, S.A., Les Molines, 2, 03450 Banyeres de Mariola, Alicante, Spain}
\affil[2]{IIIA-CSIC, Campus UAB, 08913 Cerdanyola, Catalonia, Spain}
\affil[3]{Universidad Polit\'ecnica de Valencia,  Ferr\'andiz y Carbonell, 03801, Alcoy, Alicante, Spain}

\bibliographystyle{apa}
\setcitestyle{authoryear,open={(},close={)}}
\maketitle

\begin{abstract}
Selecting the best policy to keep the balance between what a company holds in cash and what is placed in alternative investments is by no means straightforward. We here introduce PyCaMa, a Python module for multiobjective cash management based on linear programming that allows to derive optimal policies for cash management with multiple bank accounts in terms of both cost and risk of policies.
\end{abstract}

\section{Motivation and significance \label{sec:motivation}}

Cash managers usually deal with multiple banks to receive payments from customers and to send payments to suppliers. Operating such a cash management system implies a number of transactions between accounts, what is called a policy, to maintain the system in a safe state, meaning that there exists enough cash balance to face payments and avoid an overdraft. In addition, optimal policies allow to keep the sum of both transaction and holding costs at a minimum. However, cash managers may be interested not only in cost but also in the risk of policies. Hence, risk analysis can also be incorporated as an additional goal to be minimized in cash management. As a result, deriving optimal policies in terms of both cost and risk within systems with multiple bank accounts is not an easy task. To this end, we here introduce PyCaMa, a software tool to provide such optimal policies.

Despite the recent advances in cash management \cite{da2015stochastic}, there is a lack of supporting software to aid the transition from theory to practice. In order to fill this gap, we provide a cash management module in Python for practitioners interested in building decision support systems for cash management. In addition, this software allows to tackle open research questions such as: (i)~managing multiple bank accounts \cite{baccarin2009optimal}; (ii) the impact of cash flow forecasting accuracy in the cost of policies \cite{gormley2007utility,salas2017empowering}; (iii) the utility of different risk measures in cash management \cite{salas2016multi}; and (iv) robust optimization \cite{soyster1973technical,ben2002robust}.

In practice, cash management systems can be represented as a set of bank accounts and a set of transactions between them. These systems can be introduced in PyCaMa by means of an incidence matrix establishing the relationship between allowed transactions and bank accounts. Once a cash management system is defined, cash managers should describe the current cost structure including fixed and variable costs for each transaction, and holding costs for each bank account. If available, PyCaMa also accepts cash flow forecasts to reduce the uncertainty about the future \cite{stone1972use,gormley2007utility}. In addition, minimum balances for each account can be set for precautionary purposes. Finally, PyCaMa provides optimal policies for a given planning horizon by solving a linear program using a state-of-the-art mathematical programming solver such as Gurobi~\cite{gurobi}. 

Summarizing, PyCaMa is a Python-Gurobi tool aimed to automate multiobjective decision making in cash management. To the best of our knowledge, PyCaMa is the first software tool to solve the multiobjective cash management problem with multiple bank accounts. PyCaMa contributes to support cash management decision-making: (i) by empowering cash managers to derive optimal cash policies within a real-world context in which cash management systems with multiple bank accounts are the rule rather than the exception; and (ii) by providing a computational finance framework that can be used either as a tool for empirical research or as a benchmarking for further research in cash management. Next, we describe the optimization problem that PyCaMa solves.

\section{The cash management problem with multiple bank accounts\label{sec:formulation}}

\begin{figure}[htb]
\centering
\begin{tikzpicture}[bend angle=45,node distance = 1.5cm]
\node[place] (account2)  {$2$};
\node[place] (account1) [left of= account2, node distance = 4cm]  {$1$}
	edge[pre, bend left] node[above] {$x_{1,t}$}  (account2)
    edge[post] node[above] {$x_{2,t}$} (account2);
\node[place] (invest_account) [below of= account2, left of= account2]  {$3$}
	edge[pre, bend right] node[right] {$x_{4,t}$}  (account2)
    edge[post] node[left] {$x_{3,t}$} (account2)
    edge[pre, bend left] node[left] {$x_{6,t}$}  (account1)
    edge[post] node[right] {$x_{5,t}$} (account1);
\node [text centered, above of=account2] (f2) {$f_{2,t}$}
	edge[post] (account2);  
\node [text centered, above of=account1] (f1) {$f_{1,t}$}
	edge[post] (account1); 
\end{tikzpicture}
\caption{\label{fig:example} A cash management system with three accounts.}
\end{figure}
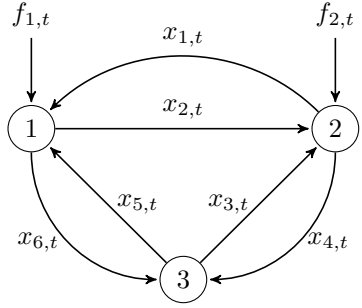

In order to formulate the problem, we first define a cash management system as a set of bank accounts and their relationship such as the one depicted in Figure \ref{fig:example}. Any cash management system with $m$ bank accounts and $n$ allowed transactions can be represented by an $n \times m$ incidence matrix $A$, with element $a_{ij}=1$ if transaction $i$ adds cash to account $j$, $a_{ij}=-1$ if transaction $i$ removes cash from account $j$, and $a_{ij}=0$ when no transaction is allowed between accounts. In the usual case of linear transaction costs between accounts with a fixed part $\gamma_0$, and a variable part $\gamma_1$, the transaction cost function $\Gamma(\boldsymbol{x}_t)$ at time $t$ is defined as:
\begin{equation}
\Gamma(\boldsymbol{x}_t) = \boldsymbol{\gamma}_0^T \cdot \boldsymbol{z}_t + \boldsymbol{\gamma}_1^T \cdot \boldsymbol{x}_t    
\label{eq:trans_cost}
\end{equation}
where $\boldsymbol{z}_t$ is an $n \times 1$ binary vector with element $z_{i}$ set to one if the $i$-th element of $\boldsymbol{x}_t$ is not null, and zero otherwise; $\boldsymbol{\gamma}_0$ is a $n \times 1$ vector of fixed transaction costs for each transaction; and $\boldsymbol{\gamma}_1$ is a $n \times 1$ vector of variable transaction costs. On the other hand, the expected holding cost function at time $t$ is usually expressed as:
\begin{equation}
H(\boldsymbol{\hat{b}}_t)= \boldsymbol{v}^T \cdot \boldsymbol{\hat{b}}_t
\label{eq:hold_cost}
\end{equation}
where $\boldsymbol{v}$ is an $m \times 1$ column vector with the $j$-th element set to the holding cost per money unit for account $j$. 

\subsection{Single objective optimization: cost}
\label{sec:cost}

Consider now a cash planning horizon of $\tau$ time steps, e.g., the next~5 working days. Within a single cost objective, given an initial cash balance $\boldsymbol{b}_0$, the solution to the problem, namely, the policy $X = \langle \boldsymbol{x}_1, \boldsymbol{x}_2, \ldots, \boldsymbol{x}_{\tau} \rangle$ that minimizes the sum of transaction and holding costs, up to the time step $\tau$, can be obtained by solving the following linear program:
\begin{equation}
\operatorname{min} \sum_{t=1}^{\tau} c(\boldsymbol{x}_t) = \sum_{t=1}^{\tau} \left( \Gamma(\boldsymbol{x}_t) + \boldsymbol{v}^T \cdot \boldsymbol{\hat{b}}_t \right)
\label{eq:lp}
\end{equation}
subject to:
\begin{equation}
\boldsymbol{\hat{b}}_{t-1} + \boldsymbol{\hat{f}}_t + A^T \boldsymbol{x}_t =  \boldsymbol{\hat{b}}_t
\label{eq:contcons}
\end{equation}
\begin{equation}
\boldsymbol{\hat{b}}_t  \geq \boldsymbol{\hat{b}}_{min} 
\label{eq:minbal}
\end{equation}
\begin{equation}
\boldsymbol{x}_t \in \mathbb{R}_{\geq 0}^{n} 
\label{eq:domain}
\end{equation}
\begin{equation}
t=1,2, \ldots, \tau
\label{eq:time}
\end{equation}
where $\boldsymbol{\hat{b}}_{t-1}$ and $\boldsymbol{\hat{b}}_t$ are $m \times 1$ vectors with previous and current balances for each account, respectively; $\boldsymbol{\hat{f}}_t$ is an $m \times 1$ vector with expected net cash flows for each account; and finally, $\boldsymbol{x}_t$ is an $n \times 1$ vector with the set of transactions (control actions) occurred at time $t$; and $\boldsymbol{\hat{b}}_{min}$ is a $m \times 1$ vector of minimum cash balances.

\subsection{Multiobjective optimization: cost and risk}
\label{sec:costrisk}

However, cash managers may also be interested in the risk of cash management policies \cite{salas2016multi}. Similarly to the definition of Conditional Value-at-Risk in \cite{rockafellar2002conditional}, we consider the Conditional Cost-at-Risk ($CCaR$) measure of policy $X$, which we define as the conditional excess expectation above a particular cost reference $c_0$ as follows: 
\begin{equation}
CCaR(X,c) = E[c(\boldsymbol{x}_t)|c(\boldsymbol{x}_t)>c_0], \hspace{2mm} \forall \boldsymbol{x}_t \in X.
\end{equation}
An additional advantage of $CCaR$ is that it is a coherent measure of risk in the sense of \cite{artzner1999coherent}. Minimizing $CCaR$ is equivalent to minimize the sum of the positive deviation of cost above reference $c_0$. As a result, we next incorporate $CCaR$ as an additional goal to be optimized through the following multiobjective linear program in which cost-risk preferences are expressed by means of weights $w_1$ and $w_2$:

\begin{equation}
\operatorname{min} \hspace{2mm} \left[ \frac{w_1}{C_{max}}\sum_{t=1}^n c(\boldsymbol{x}_t) + \frac{w_2}{R_{max}}\sum_{t=1}^n \delta_t^+ \right]
\label{eq:lp_model2}
\end{equation}
subject to:
\begin{equation}
\boldsymbol{\hat{b}}_{t-1} + \boldsymbol{\hat{f}}_t + A^T \boldsymbol{x}_t =  \boldsymbol{\hat{b}}_t
\label{eq:contcons2}
\end{equation}
\begin{equation}
\boldsymbol{\hat{b}}_t  \geq \boldsymbol{\hat{b}}_{min} 
\label{eq:minbal2}
\end{equation}
\begin{equation}
c(\boldsymbol{x}_t)  - \delta_t^+ \leq c_0
\label{eq:devcon2}
\end{equation}
\begin{equation}
\sum_{t=1}^n c(\boldsymbol{x}_t) \leq  C_{max}
\label{eq:budget1}
\end{equation}
\begin{equation}
\sum_{t=1}^n \delta_t^+  \leq  R_{max}
\label{eq:budget2}
\end{equation}
\begin{equation}
\boldsymbol{x}_t \in \mathbb{R}_{\geq 0}^{n} 
\label{eq:domain2}
\end{equation}
\begin{equation}
w_1 + w_2 = 1 
\end{equation}
\begin{equation}
t=1,2, \ldots, \tau
\label{eq:time2}
\end{equation}
where $\delta_t^+$ is an auxiliary variable used to measure deviations from a cost reference as in goal programming \cite{aouni2014financial}. $C_{max}$ and $R_{max}$ can be regarded as budget limitations for both cost and risk, leading to unfeasible policies when these constraints are no satisfied.

\section{Software description \label{sec:descrip}}

Since the cash management problem is an optimization problem, PyCaMa is based on linear programming to provide optimal policies. However, it is important to highlight that the linear programs described in Section \ref{sec:cost} (only for cost) and in Section \ref{sec:costrisk} (for both cost and risk) should be considered as baseline models that can be used for benchmarking purposes.

A cash management system, a cost structure and a set of minimum balances are sufficient to create an instance of the \textit{multibank} class. It is assumed that cash managers are able to produce cash flow forecasts for each bank account as an additional input to the problem. Otherwise, forecasts must be set to zero. Next, given an initial condition and a set of cash flow forecasts for the immediate future, cash managers can derive optimal policies either in terms of only cost or in terms of both cost an risk.

\subsection{Software architecture \label{sec:arch}}

PyCaMa is organized around the \textit{multibank} class. An instance of this class is initialized by means of different data structures in Python such as: (i) a list of $m$ banks; (ii) a list of $n$ transactions; (iii) an $n \times m$ incidence matrix $A$, establishing the cash management system; (iv) two dictionaries linking transactions and both fixed ($\boldsymbol{\gamma}_0$) and variable ($\boldsymbol{\gamma}_1$) transaction costs; (v) a dictionary linking each bank account to holding costs in vector $\boldsymbol{v}$; (vi) a list with $m$ minimum cash balances in $\boldsymbol{\hat{b}}_{min}$; as follows:

\begin{lstlisting}[language=Python]
myproblem = multibank(banks, trans, A, g0, g1, v, b_min)
\end{lstlisting}  

Once a \textit{multibank} object is created, a number of methods are implemented to provide cash managers with useful functionalities that we next describe. 

\subsection{Software functionalities \label{sec:fun}}

Cash managers can retrieve the main characteristics of the cash management system they are dealing with by using function \textit{describe}. All the input data is then shown for descriptive purposes. The main functionality of PyCaMa is function \textit{solvecost}, which provides a solution (if any) for the linear program encoded by equations \eqref{eq:lp}-\eqref{eq:time}. Given a list of length $m$ with an initial condition $\boldsymbol{b}_0$, and a $\tau \times m$ matrix $F$ of forecasts (with elements set to zero if not available) obtained by concatenating vectors $\boldsymbol{\hat{f}}_t$ with $t$ ranging in $1, 2, \ldots, \tau$, the optimal policy is obtained by executing:

\begin{lstlisting}[language=Python]
solution_1 = myproblem.solvecost(b0, F)
\end{lstlisting}

If the linear program has a feasible solution, function \textit{solvecost} returns its optimal policy for each transaction and time step. Otherwise, \textit{solvecost} warns the user that it was unable to find a solution. In addition, a $\tau \times n$ matrix with the optimal policy, and a $\tau \times m$ matrix with optimal balances derived from the last optimization can be retrieved by means of functions \textit{policy} and \textit{balance}, respectively, and the last objective value by calling the attribute \textit{objval} of the \textit{multibank} object. 

Furthermore, cash managers interested in minimizing not only cost but also the risk of policies measured by the $CCaR$, can call the function \textit{solverisk}, which provides a solution (if any) for the linear program encoded by equations \eqref{eq:lp_model2}-\eqref{eq:time2}, given $b_0$ and $F$, a cost reference $c_0$, $C_{max}$ and $R_{max}$ budget limitations, and weights $w_1$ and $w_2$, by executing: 

\begin{lstlisting}[language=Python]
solution_2 = myproblem.solverisk(b0, F, c0, Cmax, Rmax, w1, w2)
\end{lstlisting}

Summarizing, the Python input and outputs of PyCaMa are shown in Table~\ref{tab:inputs}.

\begin{table}[htbp]
  \centering
  
  \caption{Python inputs and outputs of PyCaMa}
    \resizebox{\textwidth}{!}{  
    \begin{tabular}{l|lp{2.1cm}}
    \toprule
    Inputs & Outputs & Function or attribute  \\
    \midrule
    List of banks & Description of the system & \textit{describe} \\
    List of transactions & Cost optimal policy list & \textit{solvecost} \\
    Incidence matrix & Cost-Risk optimal policy list & \textit{solverisk} \\
    Dictionary of transaction costs & Last optimal policy matrix & \textit{policy} \\
    Dictionary of holding costs & Last optimal balance matrix & \textit{balance} \\
    List of minimum balances & Last objective value & \textit{objval} \\
    Matrix of forecasts &  Cost budget     & \textit{costmax}  \\
    List of initial cash balances &  Risk budget      & \textit{riskmax} \\
    Cost reference & Cost reference      & \textit{costref}  \\
    Cost and risk maximum budgets &  Cost weight     &  \textit{costweight} \\
    Cost and risk weights &  Risk weight     & \textit{riskweight} \\
    \bottomrule
    \end{tabular}}
  \label{tab:inputs}%
\end{table}%

\section{An illustrative example \label{sec:example}}

Consider again the cash management system of Figure \ref{fig:example} with two current bank accounts 1 and 2, and an investment account 3. Temporary idle cash balances can be invested in short-term marketable securities and bonds through an investment account 3 with an average return of 3.6\% per annum, equivalent to 0.01\% per day. This is equivalent to set a holding cost 0.01\% per day for both accounts 1 and 2. Transactions are allowed between all three accounts and charged with fixed ($\gamma_0$) and variable ($\gamma_1$) costs determining the cost structure detailed in Table~\ref{tab:example}.

\begin{table}[!htb]
\centering
  \caption{\label{tab:example} Cost structure data for the example.}
    \begin{tabular}{ccccc}
    \hline
    Transaction & $\gamma_0$ (\euro) & $\gamma_1$ (\%) & Account & $v$ (\%) \\
    \hline
    1 & 50   & 0      & 1     & 0.01\\
    2 & 50   & 0      & 2     & 0.01\\
    3 & 100  & 0.01      & 3     & 0 \\
    4 & 50   & 0.001  &       &  \\
    5 & 100  & 0.01  &       &  \\
    6 & 50   & 0.001  &       &  \\
    \hline
    \end{tabular}%
\end{table}%

Assume also that a hypothetical cash manager can obtain forecasts with a maximum cumulative error of 2 million euros for a planning horizon of five days. As a result, she sets a minimum cash balance of 2 million for accounts 1 and 2. Given a initial cash balance $b_0 = \left[5, 8, 12 \right]$, for accounts 1, 2 and 3, she aims to derive the optimal policy for the next five days. To this end, let us assume that she obtains the next matrix of forecasts (with figures in millions of euros) by applying some predictive method as in \cite{salas2017empowering}.
\begin{equation}
F = \left[ \begin{array}{rrrrr}
1 & -3 & 0 \\ 
1 &-9 & 0 \\
6 & 6 & 0 \\
-1 & 4 & 0 \\
-3 & 6 & 0 \\
\end{array} \right].
\label{eq:CFexample}
\end{equation}

An instance of the \textit{multibank} class is created by introducing all the required input data as follows: 

\begin{lstlisting}[language=Python]
from PyCaMa import *                          # Import module
banks = [1, 2, 3]                             # Bank accounts
trans = [1, 2, 3, 4, 5, 6]                    # Transactions
g0 = {1:50, 2:50, 3:100, 4:50, 5:100, 6:50}   # Fixed costs
g1 = {1:0, 2:0, 3:100, 4:10, 5:100, 6:10}     # Variable costs
bmin = [2, 2, 0]                              # Minimum balances
v = {1:100, 2:100, 3:0}                       # Holding costs
A = np.array([[ 1, -1,  0,  0,  1, -1],
              [-1,  1,  1, -1,  0,  0],
              [ 0,  0, -1,  1, -1,  1]]).T    # Incidence matrix
myproblem = multibank(banks, trans, A, g0, g1, v, bmin)
\end{lstlisting}

Then, cash managers can derive the optimal policy by executing function $solvecost(b_0, F)$:

\begin{lstlisting}[language=Python]
b0 = [5, 8, 12]                               # Initial balance
F = np.array([[ 1, -3, 0],                    # Forecast matrix
              [ 1, -9, 0],        
              [ 6,  6, 0], 
              [-1, -4, 0], 
              [-1,  6, 0]])
solution_1 = myproblem.solvecost(b0, F)       # Solution
\end{lstlisting}

\begin{figure}[!htb]
	\centering
	\includegraphics[width=0.9\textwidth]{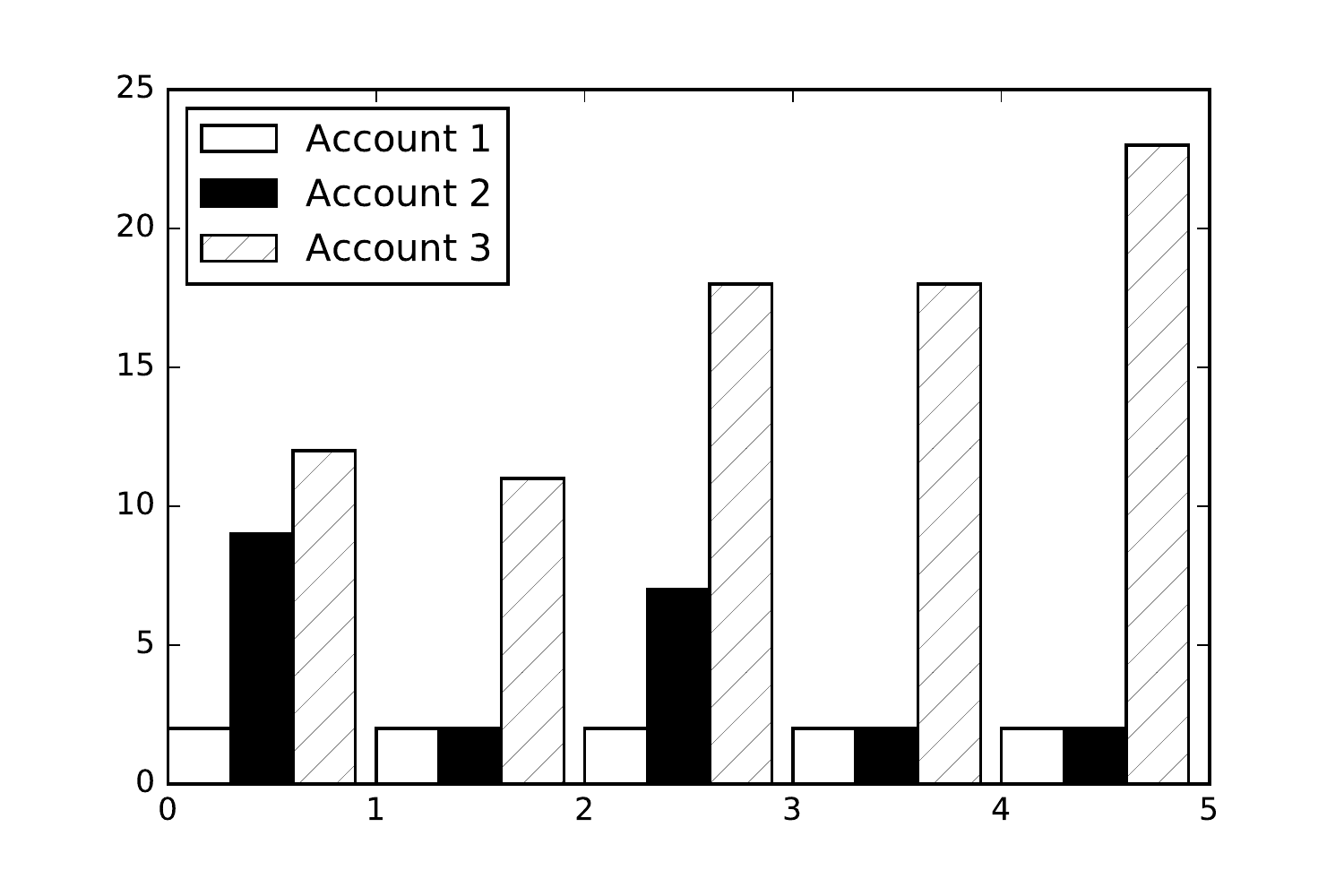}
    \caption{Optimal balances for the cost minimization example.}
	\label{fig:balances}
\end{figure}

A more compact representation of the optimal policy and balances can be obtained by calling functions \textit{policy} and \textit{balance}, which is ready to be visualized through common plotting commands in Python as shown in Figure~\ref{fig:balances}. The same instance of the \textit{multibank} class problem can now be solved in terms of cost and risk by setting a cost reference $c_0=3,000$ \euro, budget limits $C_{max}=5,000$ \euro, $R_{max}=5,000$~\euro, and weights $w_1=w_2=0.5$, and by executing $solverisk(b_0, F, c_0, C_{max}, R_{max},w_1, w_2)$, resulting in a slightly different policy.

\section{Impact \label{sec:impact}}

PyCaMa is a cash management tool that can be used either to automate decision-making in cash management or to support scientific discovery in the context of computational finance. More precisely, PyCaMa is a promising tool to tackle the following set of open research questions:

\begin{itemize}
\item The cash management problem with multiple bank accounts. There is a lack of research about multidimensional cash management systems with the exception of \cite{baccarin2009optimal}, who followed a rigorous theoretical approach. PyCaMa offers a suitable way to follow a more practical approach by providing support for research on actual-world scenarios.
\item The impact of cash flow forecasting accuracy. Although the utility of forecasts in cash management was initially demonstrated by \cite{gormley2007utility} and confirmed by \citep{salas2017empowering}, both approaches were restricted to a single bank account. PyCaMa allows to extend this analysis to cash management systems with multiple bank accounts. As a result, PyCaMa represents a tool to find the best forecasting models and their potential benefits derived from improving predictive accuracy.
\item The utility of different risk measures. A multiobjective approach to the cash management problem has been recently proposed in \cite{salas2016multi} in which the risk of alternative policies is measured by the standard deviation of daily costs. The utility of alternative risk measures can be evaluated by easily extending PyCaMa to consider different risk measures.
\item Robust optimization. Two robust approaches to optimization problems were proposed by \cite{soyster1973technical} and \cite{ben2002robust} to deal with uncertainty. PyCaMa can be used to help researchers compare existing and new robust approaches to the cash management problem in terms of both cost and risk.
\end{itemize}

Moreover, under the realistic assumption of time-varying circumstances, cash managers and researchers are allowed to modify the cost structure to analyze to what extent a change in any of the parameters of the cost structure leads to different optimal policies and, ultimately, to a variation in the cost (and risk) of managing cash. It is also important to highlight that, since \cite{miller1966model} different optimization models have been proposed based on a set of bounds. Determining such bounds may be problematic in practice due to the strong assumptions made on the probability distribution of cash flows. PyCaMa do not assume any particular form of the cash flow generating process allowing an unconstrained optimization procedure in the sense that no restriction is placed neither on the form of the policy nor on the distribution of cash flows. As a result, we expect that PyCaMa will be progressively adopted by cash managers and researchers as a more efficient tool to automate decision-making in cash management.

\section{Conclusions \label{sec:conclusions}}

In this paper, we have introduced PyCaMa, a Python module for cost and risk multiobjective optimization within a context of cash management systems with multiple bank accounts. PyCaMa solves the cash management problem when it is formulated as a linear program that aims to minimize either only cost or both cost and risk of cash policies. PyCaMa is implemented through the Gurobi Python modeling environment as a powerful and flexible way to allow an easy integration of its functionality in a more general application. Through an illustrative example, we have shown the key features of PyCaMa, and we have demonstrated how PyCaMa allows users to model complex cash management systems in an intuitive manner transforming a graphical representation in an optimization model ready to find a solution and to further experimentation. We firmly believe that PyCaMa can be a helpful tool for academic research and financial decision-support software development in the field of short-term financial planning. Natural extensions of PyCaMa include the implementation of different forecasting techniques and additional measures of risk to be added to the current functionality.

\bibliographystyle{elsarticle-num} 
\bibliography{biblio}

\end{document}